# Image guidance in deep brain stimulation surgery to treat Parkinson's disease: a review

Yiming Xiao, *Member, IEEE,* Jonathan C. Lau, Dimuthu Hemachandra, Greydon Gilmore, Ali R. Khan, *Member, IEEE,* and Terry M. Peters, *Life Fellow, IEEE*

*Abstract*—***Deep brain stimulation (DBS) is an effective therapy as an alternative to pharmaceutical treatments for Parkinson's disease (PD). Aside from factors such as instrumentation, treatment plans, and surgical protocols, the success of the procedure depends heavily on the accurate placement of the electrode within the optimal therapeutic targets while avoiding vital structures that can cause surgical complications and adverse neurologic effects. While specific surgical techniques for DBS can vary, interventional guidance with medical imaging has greatly contributed to the development, outcomes, and safety of the procedure. With rapid development in novel imaging techniques, computational methods, and surgical navigation software, as well as growing insights into the disease and mechanism of action of DBS, modern image guidance is expected to further enhance the capacity and efficacy of the procedure in treating PD. This article surveys the state-of-the-art techniques in image-guided DBS surgery to treat PD, and discusses their benefits and drawbacks, as well as future directions on the topic.***

*Index Terms*—**Deep brain stimulation, surgical navigation, neurosurgery, Parkinson's disease, image processing, MRI**

## I. Introduction

AFFECTING more than 10 million people worldwide, Parkinson's disease (PD) is a chronic and progressive neurodegenerative disorder. Although it is still primarily characterized by related motor symptoms, including tremor, muscle rigidity and bradykinesia, the associated non-motor symptoms are being increasingly recognized. In addition to pharmacotherapy, deep brain stimulation (DBS) is a surgical treatment that can improve dopamine-related motor dysfunctions of the disorder, by implanting electrodes to stimulate designated regions in the brain. Often, intra-operative micro-electrode recording (MER) is performed to confirm or refine the therapeutic target determined in the pre-surgical plan before the DBS lead is placed and secured in place. With the more commonly employed stimulation targets of the subthalamic nucleus (STN) and globus pallidus interna (GPi) to treat PD, the ventral intermediate nucleus (Vim) of the thalamus is an option for tremor-dominated PD.

In DBS surgery, the electrode must be inserted through a bur-hole to reach the optimal locus without damaging vital anatomy (e.g., blood vessels and ventricles) or stimulating other brain structures that can induce adverse responses [1]. The procedure has three main challenges. First, the DBS targets are often relatively small and poorly visualized using conventional medical imaging techniques for neurosurgical planning. Second, to avoid complications and unwanted outcomes, DBS lead insertion planning needs to consider various medical images that reveal different physiological information, which can be challenging and time-consuming for the surgeon to navigate. Finally, intra-operative tissue shift and post-operative DBS lead migration can occur, and intra-operative refinement of stimulation target is often needed. Since the inception of DBS, image guidance has played important roles to tackle these issues from three major directions, including surgical targeting, navigation, and monitoring. In targeting, many specialized MRI techniques and computational methods have been developed to reveal the location and geometry of the surgical targets, especially for the STN, and more recently, to ensure optimal functional outcomes via incorporating functional and neural connectivity data. In general, based on the approaches used, the targeting methods can be grouped into structural and functional categories. The structural category can be further organized into indirect and direct methods. For surgical navigation, multiple neuronavigation software packages, with graphical rendering of surgical data, feature integrated image-processing algorithms, and automatic lead trajectory planning have been proposed to facilitate both pre-surgical planning and post-hoc outcome analysis. Lastly, for DBS monitoring, imaging and computational methods have been pursued to ensure correct electrode placement, as well as potentially monitoring functional responses to the stimulation and hemorrhage.

Image-guidance in DBS surgery is a highly multi-faceted topic that involves image acquisition, image processing, computational models, and physiological modeling of the tissue and brain function. The advancement of DBS image guidance is imperative to ensure and possibly extend the DBS outcomes. while simultaneously offering needed insights

Y. Xiao, J.C. Lau, D. Hemachandra, A.R. Khan, and T. M. Peters are with Robarts Research Institute, Western University, London, N6A 5B7, Canada (e-mail: yxiao286@uwo.ca).

J.C. Lau is with the Department of Clinical Neurological Sciences, Division of Neurosurgery, Western University, London, Canada

D. Hemachandra, G. Gilmore, A.R. Khan, and T. M. Peters are with the School of Biomedical Engineering, Western University, London Canada.

A.R. Khan, and T. M. Peters are with the Department of Medical Biophysics, Schulich School of Medicine and Dentistry, Western University, London, Canada.

A.R. Khan is with the Brain and Mind Institute, Western University, London, Canada.

regarding our brain circuitry and the true potential of DBS, as well as providing the knowledge for other therapies (e.g., focused-ultrasound surgery). This review primarily focuses on current methodological developments of image guidance for DBS to treat PD, in the three aforementioned aspects. Since more existing methods are devoted to surgical targeting, the review of this aspect is proportionally longer than the other two. A schematic of this review is shown in Fig. 1.

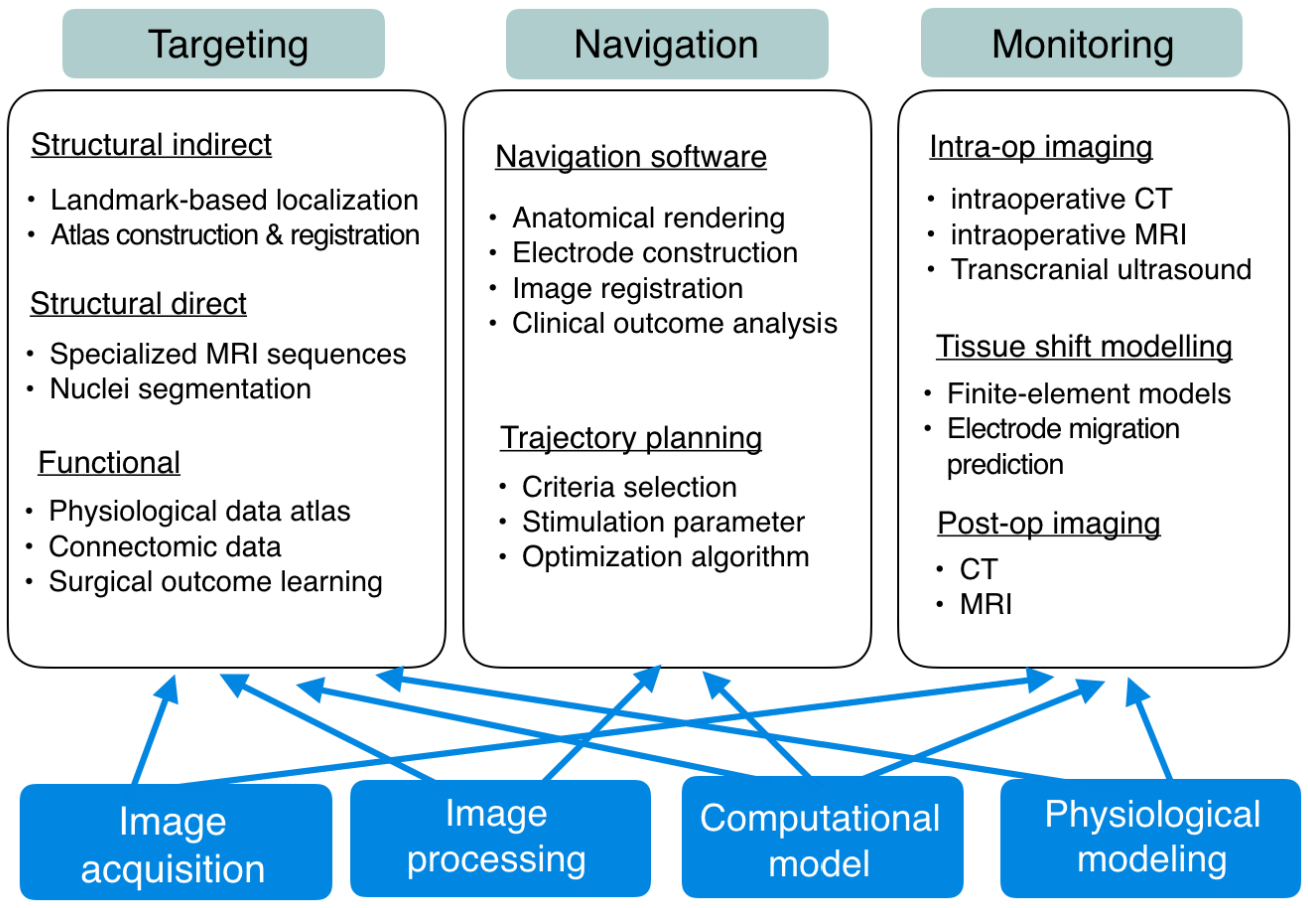

**Figure 1**. *Schematic of different components reviewed in image-guidance for DBS surgery to treat Parkinson's disease.*

## II. STRUCTURAL INDIRECT TARGETING

Indirect targeting approaches infer the centroid or geometry of the nucleus either from adjacent anatomical features, or by fitting an atlas to an individual's anatomy as represented in pre-operative scans (e.g., T1w MRI) that fail to sufficiently visualize the target. This approach was used historically before the era of modern imaging, and to this day is still often used as a first approximation of the target location.

### *A. Coordinate-based approaches*

Conventionally, the locations of the surgical targets are often inferred from their spatial coordinates in relation to more easily identifiable anatomical landmarks, as defined in well-established atlases such as those from Talairach [2] or Schaltenbrand atlases [3]. Among different coordinate systems, the mid-point (i.e., middle commissure or MC) between the line connecting anterior commissure (AC) and posterior commissure (PC) (see *Fig.2*) is the most popular landmark to help locate the surgical targets, such as the STN. Efforts were therefore, made to improve the accuracy, consistency, and automation of AC and PC identification [4, 5] through multi-template registration and local image feature learning as well as developing open standards for placing these features [6]. To adjust for individual variability, the coordinates are usually scaled by a factor that normalizes AC-PC line lengths between the patient's anatomy and the atlas employed, and some also use additional adjustments based on considerations, such as the third ventricle width [7], but there remains a lack of consensus among the specialists [8].

### *B. Structural brain atlases and image registration*

Distinct from coordinate-based targeting that only offers points in 3D, full geometry and richer anatomical context can be obtained for the surgical target, by mapping structural brain atlases to the patient's anatomy, using volume-to-volume image alignment. Besides the more recent digitized Talairach [2] or Schaltenbrand atlases [3], a number of brain atlases have been released to benefit DBS planning. As a starter, several subcortical atlases have been developed from 3D reconstruction of histological annotations [9-11], and were co-registered to single-subject T1w MRI templates. For better anatomical representation, atlases averaged from a group of subjects have become a standard practice. In 2013, Haegelen *et al.* [12] built a PD-population-averaged atlas (named ParkMedAtlas) with manual segmentation of 7 pairs of subcortical structures. Xiao *et al.* [13, 14] constructed the MNI PD25 atlas with a novel T1-T2*-fusion contrast template and a co-registered histological atlas with 123 structure labels, and the ultrahigh-resolution BigBrain atlas [15, 16] (*Fig.3a*). More recently, Inglesias et al. [17] published a probabilistic atlas of the human thalamic nuclei with *ex vivo* MRI and histology of 6 elderly subjects, and the new CIT168 atlas (*Fig.3b&c*) was created from 168 young healthy subjects by Pauli *et al.* [18]. Benefiting from enhanced tissue contrast of ultra-high field 7T MRI, Keuken *et al.* [19] and Wang *et al.* [20] proposed multi-contrast averaged brain atlases with probabilistic maps of the basal ganglia structures using healthy subjects, which was applied prospectively in surgical cases [21]. Besides anatomy, atlases that include functional sub-divisions of target nuclei were also proposed, intended to better pin-point the therapeutic subregion. In this vein, Silva *et al.* [22] and Accolla *et al.* [23] provided parcellation for the GPi and STN, respectively. Based on the atlas of Chakravarty *et al.* [11], the DISTIL atlas [24] sub-divides the STN and GPi into 3 different functional zones in the MNI152 template space.

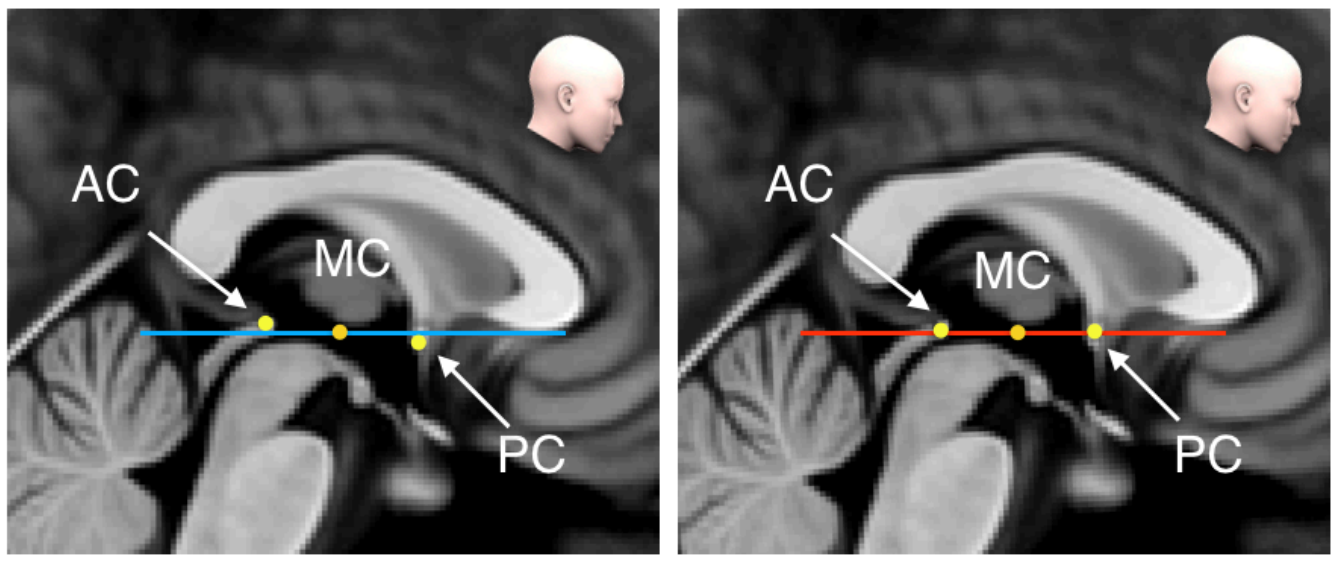

**Figure 2**. *AC-PC lines and MC points defined in Talairach atlas space (Left) and Schaltenbrand atlas space (Right), shown in the mid-sagittal plane of the T1w ICBM152 template. Note that the differences in the definition of AC-PC, and thus the reference coordinates to locate surgical targets.*

Atlas-to-MRI registration strategies can impact the quality of targeting. In indirect targeting, such registration relies only on T1w MRI. Bardinet *et al.* [25] used the patient's original and mirrored MRI to improve the atlas registration accuracy. Duay *et al.* [26] proposed an active-contour-based atlas registration for STN targeting. Using the Colin27 template [27] as a medium, Chakravarty *et al.* [28] employed pseudo-MR images derived from digitized histological atlases to map the atlas to individual patients. Finally, different deformation models and publicly available registration software were assessed by Chakravarty et al. [29] and Ewert *et al*. [30], confirming the need for refined nonlinear warping.

## III. Structural Direct Targeting

Direct targeting locates the surgical targets using MRI techniques and the associated image processing methods that can directly visualize them. Among these, 2D fast spin-echo (FSE) T2w MRI has the longest history in the clinic. However, such images are slow to acquire, making it difficult for the patients to tolerate. Often highly anisotropic resolution (3~5 mm axial slice thickness) images, or just a slab of the brain is obtained. Newer MRI sequences primarily aim to enhance the contrast of the nuclei, image quality (e.g., signal-to-noise ratio and image resolution), or scanning efficiency. Full summaries of MRI sequences and nuclei segmentations are included in *Tables S1 & S2* of the supplementary material.

### *A. Inversion recovery imaging*

Inversion recovery (IR) imaging that reduces signals from designated tissues can boost the contrasts of subcortical structures. Ishimori *et al*. [31] proposed a 3D phase sensitive IR sequence with image contrast optimization to visualize the STN. Sudhyadhom *et al.* [32] proposed the Fast Gray matter Acquisition T1 Inversion Recovery (FGATIR) image to highlight the boundaries of subcortical structures and improve atlas fitting, demonstrating that FGATIR is superior to T2-FLAIR in providing contrast between nuclei. Kitajima *et al.* [33] compared the FSE T2w and fast short inversion time IR (FSTIR) images, and concluded that it is easier to differentiate the STN from the adjacent SN in FSTIR images. Later, Nowacki *et al.* [34] evaluated the application of the modified driven equilibrium Fourier transform (MDEFT) sequence for depicting the GPi, though it was shown that magnetization transfer (MT) maps can better visualize the GP than MDEFT [35].

### *B. Susceptibility-weighted imaging*

Susceptibility-weighted imaging (SWI) enhances the contrast of the subcortical structures by taking advantage of their rich iron deposition. A typical SWI sequence produces a T2*w MRI, a phase image, and a magnitude-phase-fusion contrast. While the magnitude-phase-fusion image was originally intended for venography, the sequence has subsequently been customized to visualize the STN for DBS. Vertinsky et al. [36] optimized the scanning parameters of single-echo SWI images to best visualize the STN, and reported that the phase image is the best for the purpose. T2*w MRI exhibit high contrast for subcortical structures, but also exhibits evident susceptibility artifacts, which requires additional T1w MRI in surgical planning and may affect image registration quality. Xiao *et al.* [37] proposed a 10-echo FLASH sequence that produces different contrasts, including T1w and T2*w MRIs, whose image quality was enhanced through averaging adjacent echoes and optimizing scanning parameters. Alternatively, Volz *et al.* [38] attempted to recover the signal loss from susceptibility artifacts, by estimating pixel-wise signal loss and adaptively fusing multi-echo data. Finally, Wu *et al*. [39] proposed an inverse double-echo steady-state (iDESS) technique to depict the midbrain nuclei and reduce susceptibility signal loss.

### *C. Quantitative imaging*

Quantitative MRI techniques can reveal the microstructural architectures of brain tissues by deriving their intrinsic MRI properties, including T1, T2, T2*, susceptibility, and magnetization transfer (MT) parameters. Therefore, these quantitative maps offer the opportunity to better describe the boundary between adjacent tissue types. Guo *et al*. [40] introduced the driven equilibrium single pulse observation of T1 and T2 (DESPOT1 and DESPOT2 [41], respectively) for DBS implantation. Helms et al. [35] proposed a novel, semi-quantitative magnetization transfer (MT) imaging to improve contrast of basal ganglia structures, and discussed its feasibility in anatomical localization for DBS. Aside from the standard contrasts from the SWI sequence, T2* or R2* (1/T2*) maps [42] and later quantitative susceptibility maps (QSMs) [43] can also be derived to help improve the visualization of the STN and GP.

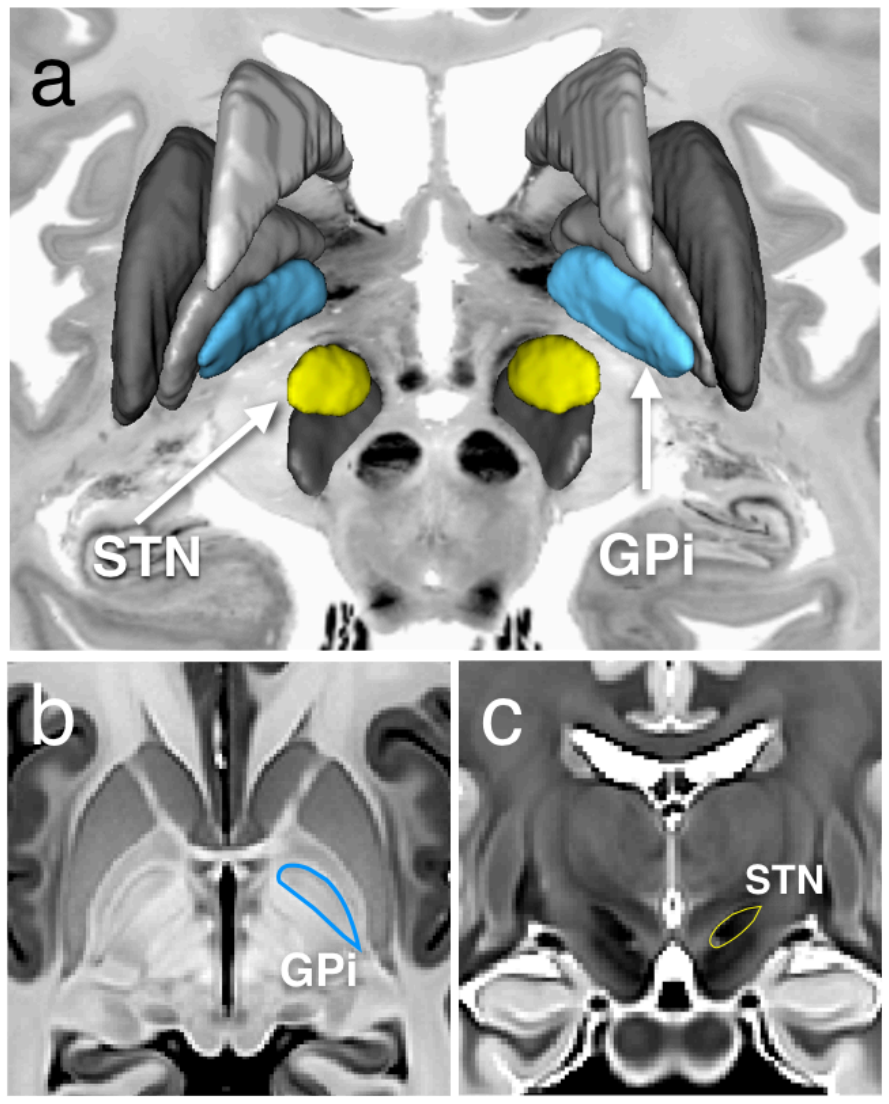

**Figure 3**. *(a) The STN and GPi rendered in 3D with other basal ganglia structures in the histological BigBrain atlas [15, 16]; (b) GPi is shown in axial view of the T1w CIT168 atlas [18]; (c) STN is shown in the coronal view of the T2w CIT168 atlas.*

### *D. Ultra-high field MR imaging*

Ultra-high field (>3T) MRI can boost the sensitivity, resolution, and signal-to-noise ratio of functional and anatomical images, and thus can offer more exquisite anatomical details. Abosch *et al.* [44] obtained T2w and SWI scans at 7T, and showcased the superior ability to effectively improve the identification of the GPi, STN, and Vim. Duchin *et al.* [45] found that T2w MRI at 7T has minimal distortion within the central part of the brain compared with at 1.5T, and thus is suitable for clinical use. Cho *et al.* [46] compared T2*w MRI at 1.5, 3, and 7T for visualizing the STN, while Deistung *et al.* [43] compared the contrast of subcortical structures in QSM, T2*w, phase, and R2* images at 7T, and concluded that QSM demonstrates the highest anatomical detail.

### E. *Deep brain nuclei segmentation*

These aforementioned MRI techniques provide the opportunity for automatic nucleus segmentation to facilitate computer-assisted DBS planning and pathological studies of PD. Thus far, a number of groups have proposed and validated their methods, which can be categorized into atlas-based (single-atlas propagation and multi-atlas label fusion), model-based, and deep learning-based methods. Distinct from atlas-based segmentation in indirect targeting, the methods in this section involve image contrasts that display the nuclei of interest with sufficient contrast for registration or atlas selection. With a larger volume and visibility in T1w MRIs, segmentation of the GPi is more commonly reported. As for STN segmentation, Haegelen *et al.* [12] compared methods including single atlas propagation with different registration algorithms, ANIMAL [47] and SyN [48], and a non-local means label-fusion method [49] to segment subcortical regions using high-resolution sectional T2w FSE MRI at 3T. Their comparison offered the best results with a 0.64 Dice score using the ANIMAL algorithm. Also with 3T data, Xiao *et al.* [50] employed T2w pseudo-MRI to improve histological atlas registration to segment the STN. Later, to resolve the drawbacks of single-atlas propagation [51], the same group proposed label fusion segmentation methods for the midbrain nuclei, using fuzzy majority-voting on T2w MRI [51] and multi-contrast non-local means on SWI data [52], with the Dice score ranging 0.74~0.90 for their approaches. With respect to model-based approaches, Li *et al.* [53] used a band-limited level set to segment the STN on 3T T2w MRI. At 7T, Visser *et al.* [54] used both shape and multi-spectral intensity (QSM, T2w, T2*w, and fractional anisotropy map) modeling to segment the midbrain nuclei, while Kim *et al.* [55] used regression models to identify the STN in 1.5T T2w MRI based on a 7T training set. Most recently, deep learning has been employed to identify the STN [56], achieving a Dice coefficient of 0.90. Intending to serve a wider community, Manjon *et al.* [57] proposed *pBrain*, an online processing pipeline for segmenting the STN based on multi-atlas label-fusion with multiple -scales and features in T2w MRI.

## IV. TARGETING WITH FUNCTIONAL DATA

Distinct from the previous two categories that focus on providing the centroid or geometry for the anatomy to be stimulated, targeting with functional data is centered around the concept that targeting a specific functional region within the nucleus of interest may provide the maximum symptom improvements while minimizing adverse effects. Three general categories of functional targeting have been reported: probabilistic functional atlas mapping; connectivity-based targeting; and machine learning for surgical outcome. A summary of all currently available methods is available in *Table S3* in the supplementary material.

Among the various approaches, the probabilistic functional atlas mapping has the longest history. For this type of atlas, the locations of the active electrodes, the responses to stimulation, and the stimulator settings, are recorded for each individual patient. The stimulation locations of a cohort of patients with sufficient clinical benefits were deformed to a common anatomical atlas, while the influence of the applied electric field was modeled by a kernel function. Finally, the results were averaged to form a probabilistic representation of the "hot spot" to guide loci selection. As the earliest to develop such atlases, Nowinski *et al.* [58, 59] showed that DBS planning with their probabilistic functional atlas is better than solely relying on an anatomical atlases. Guo *et al.* [60] also presented a probabilistic atlas, built from intra-operative data of Finnis *et al.* [61], and demonstrated the advantage of the functional atlas in comparison to other anatomy-based approaches [40]. To further improve atlas-to-patient registration for stimulation target selection, D'Haese *et al.* [62, 63] proposed a multi-template technique for their electrophysiological atlas, and reported < 1.75mm mean deviation from final implanted location when using the atlas for planning. From the same group, Pallavaram *et al.* [64] used spherical kernels instead of more commonly seen Gaussian kernels [60] to more accurately reflect the effect of electric stimulation.

The second category of techniques utilize brain connectivity data, derived from diffusion and functional MRI to locate therapeutic neural pathways and sub-regions of nuclei that are difficult to visualize in structural MRI. Tractography obtained from diffusion MRI has been employed to determine optimal electrode placement within sub-regions of the STN [65, 66] and GPi [67]. Since the Vim nucleus of the thalamus is difficult to visualize on structural images, both tractography [68] and task-based functional MRI [69] have been used to help locate it. More specifically, Sammartino *et al.* [68] reported a mean discrepancy of 1.6mm between targeting using tractography and actual surgical choice. The extraction of dentatorubrothalamic (DRT) tract using tractography by obtaining tracts passing the dentate nucleus, thalamus, and motor cortex has been proposed as an alternative image-based target for tremor [70]. Compared with the first category of functional targeting that relies on the collection of previous target locations, connectivity-based targeting, often referred to as connectomic DBS, links the therapeutic benefits with activation of specific neural pathways.

Lastly, the new data-driven approaches rely on machine learning to predict surgical outcomes at intended stimulation locations. Horn *et al.* [71] used brain connectivity data and linear regression models to calculate the surgical outcomes with an average 15% error margin for Unified Parkinson's Disease Rating Scale (UPDRS) motor score improvements. Later, Baumgarten *et al.* [72] trained an artificial neural network model with 3D coordinates and levels of electric current to predict whether selected stimulation loci induce therapeutic benefits and side effects, obtaining the sensitivity of 93.07% and 73.47%, respectively. Similarly, Lin *et al.* [73] newly proposed a random forest classifier to differentiate ineffective vs. effective targets, but using instead a mean fractional anisotropy and tractography streamline fraction of the stimulation site within the STN to the rest of the brain. Their approach achieved a classification accuracy of 84.9%.

## V. DBS NAVIGATION SOFTWARE & TRAJECTORY PLANNING

Surgical navigation software provides the needed interface to visualize complex clinical information, pre-process imaging data, and possibly devise surgical plans, as well as perform post-hoc outcome analysis. Aside from commercial software, a number of DBS navigation software platforms have been developed for research purposes. The summary of the various current research-based navigation systems is listed in *Table S4* of the supplementary material.

As the first of its kind, CranialVault [74] packaged automatic image processing tools, data visualization, and functional target selection [18] into one user interface. PyDBS [75] and IBIS NeuroNav [76] also offer surgical data visualization for planning DBS. With the first containing streamlined workflow to segment and register multi-modal scans, both of them incorporate efficient and automatic trajectory planning tools [77-80] as software plug-in extensions. Primarily to help analyze post-surgical impacts, PaCER [81] and DBSproc [82] were built to more accurately reconstruct the electrodes and offer tractographic analysis of surgical targets. On the similar note, LeadDBS [83] provides a platform to conduct statistical analysis for DBS procedures with integrated registration tools, an electrode reconstruction function, and integrated brain atlases, including both structural and connectomic types.

A safe trajectory for DBS electrode insertion should consider multiple criteria, including but not limited to the distance to the optimal stimulation target while avoiding cerebral vasculature, ventricles, and structures, such as posterior limb of the internal capsule, that can induce adverse effects. However, proposing viable insertion paths by simultaneously considering all these conditions may impose a heavy cognitive demand for the surgeon, making the process arduous and time-consuming. Therefore, several automated trajectory planning algorithms have been proposed by framing the task as optimizing cost functions that represent these criteria mathematically.

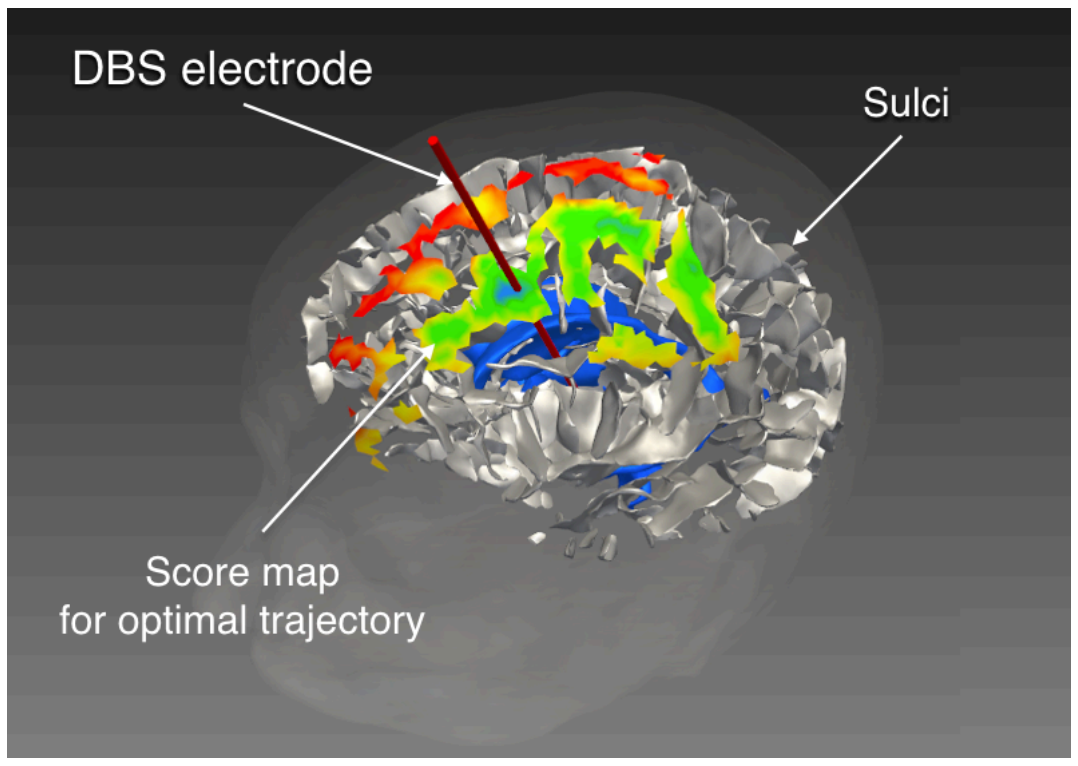

**Figure 3**. *Demonstration of the automatic DBS trajectory planning software proposed by Essert et al. [79, 80] in a case of STN stimulation (Courtesy of Dr. Caroline Essert, University of Strasbourg)*

While the earlier versions [84, 85] of these algorithms demonstrated the feasibility of such optimization frameworks with limited criteria, more recent approaches [77-80] incorporated a larger set of constraints, as well as introduced new data types into the optimization procedure. From the Montreal Neurological Institute group, Bériault *et al.* [77, 78] proposed automated DBS planning using SWI venography to ensure surgical safety and introduced a computational model for STN DBS stimulation with multiple active electrode contacts. In a similar vein, Essert *et al.* [80] emphasized the geometry of the insertion paths in their algorithm, and assessed the performance retrospectively. Later, Dergchyova *et al.* [79] further developed the Essert system, allowing the cost function to balance both the functional improvement predicted by probabilistic efficacy maps and the surgical risks. Despite various successes in automated trajectory planning algorithms, the strategy to weigh various path planning criteria is still largely determined by the users in a trial-and-error manner, and no systematic studies have been conducted to provide a unified framework to determine the optimal weights.

## VI. INTRA-OPERATIVE AND POST-OPERATIVE MONITORING

Intra-operative monitoring can ensure the safety and quality of DBS, and the technical advancements have made it feasible to inspect many crucial factors during surgery, including electrode position, brain shift, intra-operative hemorrhage, and functional responses to stimulation. This is especially helpful for brain shift in DBS, which is able to impact surgical quality [86-88] in both intra-operative targeting and post-surgical electrode migration, with an averaged electrode shift of 1.41 mm measured by Sillay *et al.* [86]. So far, besides more conventional and costly intra-operative MRI (iMRI) [89] and intra-operative CT (iCT) systems, transcranial sonography (TCS) [90, 91] and bio-mechanical modeling [87, 92] have also been reported. iCT is typically used to confirm the electrode position, and while it is an accurate equivalent to post-operative MRI measurement [88], it lacks soft tissue contrast. Besides electrode position verification, iMRI [89] can provide different image contrasts that visualize the surgical target, brain shift, and hemorrhage, and perform intra-operative fMRI [93] to possibly refine electrode placement and stimulation parameters. A more cost effective modality, 3D TCS was tested experimentally to guide DBS electrode placement [91]. With a limited accuracy of electrode tip localization [90] due to challenges in TCS-MRI fusion and image distortion, the method is not yet ready for clinical use. Lastly, finite element models (FEMs) [87, 92] have also been demonstrated in simulation [87] and clinical case study [92] to improve the electrode placement by accounting for brain shift, but further developments that allow efficient integration into the clinical workflow are needed.

Post-operative imaging is another crucial step to verify final electrode position, particularly for potential electrode migration, assessment of surgical complications, and investigation of physiological impacts of the treatment. Recent studies [94, 95] with post-operative structural and functional MRI, have revealed the tissue and functional changes after DBS surgery. While CT avoids the need for meticulous estimation of heat dissipation with MRI in the presence of a DBS electrode, the physiological insights offered is limited. Studies [96] have recommended a specific absorption rate (SAR) below 0.1 W/kg when planning post-operative MRI for DBS cases.

## VII. Discussion

### A. Targeting: coordinates, anatomy, and function

Atlas coordinate-based DBS targeting has a long history in the clinic. It assumes that the relative positions and sizes of DBS surgical targets, such as the STN, are consistent across subjects, and can tolerate minimum or lower quality imaging data. Meanwhile, non-negligible anatomical variability across subjects [51, 97, 98] has also been reported for the surgical target. The choice between the coordinate-based technique and direct targeting has been actively debated, largely for localizing the STN. While the earlier reports favored the first [99], more recent evidence [51, 98] prefer the latter. This discrepancy may likely be attributed to the improved image quality of MRI scanners, especially in terms of signal-to-noise ratio, image resolution, and newer scanning sequences, but unfortunately most studies [99] only reported results with fairly limited patients, or even healthy cohorts. Notably, the initial stimulation locations derived from coordinate-based methods and anatomical MRI are usually the estimated centroid or geometric center of the nucleus, which is often different from the therapeutic target. Inputs from functional information at the planning stage can therefore be important. Such insights can be made available through brain atlases that contain connectivity-based sub-divisions of the nucleus [24], probabilistic representation of effective stimulation loci [60, 64], and group-averaged connectomes [71]. With physiological recording [100], we have observed that therapeutic regions are not necessarily bounded by the borders of a particular anatomy. To push this notion further, functional targeting with brain connectivity data, including diffusion and functional MRI, allows effective and tailored stimulation of nuclei subregions [101] or instead towards white matter pathways [70, 102] for targeted symptoms.

Both indirect and direct targeting methods, and even many functional targeting approaches, have relied heavily on brain atlases, where significant progress has been made in adding super-high-resolution histological data [16], multi-contrast MRI [9, 13, 20, 64, 97], unbiased anatomical representation, and brain connectivity information [67, 71]. To accurately transfer the rich information from the atlases to an individual's anatomy, suitable registration strategies are crucial. As anatomical variability in midbrain nuclei was revealed by high-quality structural MRI [51, 97], refined non-linear registration strategies [29, 30] are needed, and multi-contrast registration [16, 50, 71] is recommended for targets that are not visible on T1w MRI. Aside from brain atlas customization, concerns [103] have also been raised regarding potential insufficient representation of individual physiology when using group-averaged connectomic atlases for DBS planning. Comprehensive investigation is still required to confirm the impact of these approaches.

Selecting the appropriate ground truth to assess DBS targeting methods is crucial and challenging. Typical references [99] to validate DBS planning techniques include final implantation locations shown in post-operative scans, intra-operative recording locations, coordinates shown in established atlases, structural MRI, and post-surgical symptom improvement. With possible tissue shift and electrode migration, stimulation targets determined during and after surgery may not overlap, and they may also exhibit an inherent bias away from the pre-surgical plans. Although symptom improvement appears more meaningful as a ground truth, the possibility of alternative candidate targets with better outcomes make it difficult to assess prospective cases. Surprisingly, some functional targeting approaches [59, 69] used atlas coordinates as ground truth, which is not ideal, and the differences in validation metrics make it difficult to conduct cross-method comparisons. Finally, most reported planning techniques focus only on single contact stimulation. Multi-contact DBS planning, which is more complex to validate, is seldom mentioned in the literature.

### B. MR imaging and segmentation of surgical targets

To date, a majority of the literature on novel structural MRI sequences for DBS is dedicated to the STN, for which, susceptibility-based contrasts [36, 37, 42, 44, 46], particularly 7T QSM [43] showed superior results. Most MRI techniques surveyed in *Table S1* were only developed based on small cohorts with primarily healthy subjects. This may affect their performance in PD patients, who can exhibit varied biochemical features in the brain and are less tolerant to relatively long scan sessions. In these novel sequences, typical slice thickness is still *2 mm*, which is not ideal to accurately depict the geometry of the small STN, and is also in contrast to the finer isotropic resolutions used in automatic segmentations in *Table S2*. Unfortunately, none of these newer sequences has been reported in routine surgical practice in the literature. Another important factor of MR imaging for DBS is geometric distortion, which can originate from various sources, such as magnetic field inhomogeneity, chemical shift, and susceptibility differences. This factor was only assessed by Ishmori *et al.* [31] and Duchin *et al.* [45] in their studies, without disentangling individual contributing sources. With ultra-high field 7T MRI becoming more available, the inherent issues of image distortion from increasing field strength, particularly magnetic field inhomogeneity, will require rigorous examination [104].

Automatic segmentation algorithms for the STN have provided excellent results, achieving a Dice score of nearly 0.90. Unlike other brain structures that can be easily visualized in standard T1w MRI, the technical developments for segmenting DBS targets, especially deep learning algorithms, were encumbered by limited quality MRI databases, the lack of the unified segmentation protocols, and effective MRI technique development. Yet, the existing techniques may not offer the same accuracy as reported on clinical scans, which often exhibit poorer rougher image quality.

Although the benefits of connectomic DBS surgery are becoming increasingly evident, especially for targets that are difficult to depict solely by structural imaging, high quality diffusion and functional MRI face challenges in their integration into clinical routines. This makes development and validation of relevant functional targeting algorithms with large clinical databases more challenging.

### C. *Intra-operative and post-operative imaging in DBS*

Surgical target shift in DBS due intraoperative tissue shift has been measured to be more than 2mm in some cases [92, 105, 106]. However, aside from surgical protocols that avoid CSF leakage to reduce this phenomenon, no effective algorithms similar to those in brain tumor resection [107] have been proposed based on inter-operative imaging for tissue shift correction. Intra-operative fMRI [93] offers the possibility of real-time monitoring for the impacts of DBS to improve the final targeting, but the high cost and special requirements for surgical setup make it difficult to access. Aside from confirming electrode position, post-operative MRI can also play a valuable role in advancing our understanding of the mechanism and true impacts of DBS [94, 95]. For both intra-operative and post-operative imaging, image artifacts (e.g., streaking and distortion) induced from DBS leads can pose difficulty in obtaining accurate knowledge of final implantation location, and need to be carefully considered in imaging and outcome analysis.

### D. *Future directions*

Studies [108] have indicated that instead of local circuitry modulation, DBS may exhibit widespread effects on the brain. Although still at the early stage, preliminary retrospective studies in connectomic DBS [70, 71, 73] have shown great potentials. The surgical strategy may eliminate or at least minimize the need for MER, potentially resulting in shorter operation time and improved patient comfort. The use of brain connectivity information in DBS planning and analysis will continue to enrich our understanding of the neural circuitry and the mechanism of DBS in the future.

Current DBS strategy focuses primarily on dopamine-related motor symptoms of PD, and the non-motor issues (e.g., psychiatric symptoms and cognitive declines) are rarely considered. In addition, traditional stimulation targets, such as STN and GPi still have their limitations in therapeutic benefits and side effects [109]. These motivate the search for new DBS targets to treat PD, and a number of potential candidates, including individual nuclei and fiber pathways, have been proposed [70, 110], as well as multi-target stimulation to boost the impacts [111]. These demands pose two challenges on image-guidance of DBS. First, continued efforts are still needed to develop new MRI (structural and functional) techniques and the associated analysis methods to help identify the structures/pathways and confirm their impacts. Second, with richer image modalities required to improve DBS strategy, the challenges of multi-target stimulation, and new DBS devices, DBS navigation software will need to provide more intuitive and resource-efficient strategies for data visualization [112] and updated algorithms for improved electrode trajectory planning.

Machine learning and deep learning techniques have shown early success in facilitating DBS planning in anatomical segmentation [53, 55-57], stimulation target selection [72, 73], and treatment outcome prediction [71]. While these techniques will continue to develop in the future to benefit the clinic, most of them have relied on limited research-grade MRI scans, whose resolution and image quality are superior to clinical images. Recent developments in deep-learning-based fast MRI [113], image denoising [114], and super resolution [115], along with hardware improvements, are expected to facilitate improved clinical data acquisition. Together with the open data initiatives that enable growing publicly available imaging repositories, learning-based methods may play a more important role in the future.

The development of image-guidance techniques can also benefit other treatments for PD, such as focused ultrasound (FUS) [116]. In addition, it can also greatly contribute to asleep DBS surgery [117], where the electrode is implanted under general anesthesia, to improve the patient's comfort.

## VIII. CONCLUSION

This review provides the state of the art for medical image guidance in targeting, navigation, and monitoring of the DBS procedure to treat Parkinson's disease. With an increasing demand for more enhanced and personalized treatments, future developments in DBS image guidance are expected to focus on incorporating connectomic data for improved functional targeting, updating imaging and neuronavigation techniques for new brain stimulation strategies (e.g., novel targets and multi-target stimulation), and finally, leveraging machine/deep learning to allow translation of the knowledge and tools developed in research for clinical data and workflows. Historically, image guidance in DBS has closely accompanied the birth and evolvement of this procedure, and will continue to contribute to its future development by providing in-depth insights of its mechanism and possibly extending its clinical benefits.

## ACKNOWLEDGMENT

The work is supported by the BrainsCAN and CIHR fellowships for Y. Xiao. Support from CIHR Foundation grant FDN 201409 is also acknowledged. JC Lau is funded through the Western University Clinical Investigator Program accredited by the Royal College of Physicians and Surgeons of Canada and a CIHR Frederick Banting and Charles Best Canada Graduate Doctoral Award Scholarship.

**Article title: Image guidance in deep brain stimulation surgery to treat Parkinson's disease: a review**
**Authors: Yiming Xiao, Johnathan C Lau, Dimuthu Hemachandra, Greydon Gilmore, Ali R. Khan, and Terry M. Peters**

**Table S1**. Specialized MRI sequences for visualizing surgical targets of deep brain stimulation.

| **Authors** | **Field strength** | **No. of subjects** | **MRI sequence** | **Anatomical structure** | **Resolution** | **Imaging time** | **Evaluation** | **Conclusion** |
|---|---|---|---|---|---|---|---|---|
| Ishmori et al. (2007) | 1.5T | 3 healthy & 2 patients | 3D phase-sensitive inversion recovery | STN | 1×1×2 $mm^3$ | 16:56 min | Less than 1% geometric distortion; Measured STN coordinates agree with the landmark-based approach | The sequence offers distortion-free images for DBS targeting |
| Kitajima et al. (2008) | 3T | 4 healthy & 20 patients | FSTIR | STN | 0.39×0.625×2.5 $mm^3$ | 5:25 min | Contrast ratio between STN and adjacent tissues: 0.10~0.23 | FSTIR is superior than T2w FSE in the STN targeting |
| Sudhyadhom et al. (2009) | 3T | 3 patients | FGATIR | GPi, STN, Vim | 0.8×0.8×1 $mm^3$ | 11:14 min | CNRs of STN, GPi and Vim against surrounding tissues: 5.96, 6.06 & 7.27 | FGATIR reveals richer anatomical features than conventional T1w and T2w MRI |
| Vertinsky et al. (2009) | 3T | 8 healthy | SWI phase map using SENSE | STN | 0.55×0.75×1.5 $mm^3$ | < 2:30 min | Best visual scores based on 3 raters with TE=20ms and SENSE=1.40 | Delineation of the STN was superior in phase images |
| Volz et al. (2009) | 3T | 4 healthy & 4 patients | T2*w multi-echo FLASH | STN | 1×1×2 $mm^3$ | 3:30 min | Mean CNR of STN vs. WM: 37.1 | Combining multi-echo information can reduce susceptibility artifacts and improves the STN contrast |
| Abosch et al. (2010) | 7T | 6 healthy | SWI | GPi, STN & Vim | 0.95×0.95×0.95 $mm^3$ | 15:00 min | Qualitative visual comparison promotes SWI over T2w MRI in subcortical structure visualization | MRI at 7 T have yielded improved anatomic resolution of deep brain structures |

Supplementary material

| | | | | | | | | |
|---|---|---|---|---|---|---|---|---|
| Cho et al. (2010) | 7T | 11 healthy & 1 patient | 2D GRE (T2*w) | STN, GPi | $0.25\times0.25\times2$ mm$^3$ | 5:54 min | Mean normalized intensity difference of STN vs. SN: ~0.20; only visual assessments for GPi | 7T MRI images showed marked improvements in visualization STN and GPi than 3T |
| Y. Xiao et al. (2012) | 3T | 4 healthy & 2 patients | Multi-contrast Multi-echo FLASH sequence (R2* & SWI) | STN | $0.95\times0.95\times0.95$ mm$^3$ | 7:05 min | Contrast ratio between STN and SN: 0.31 (R2*map) 0.36 (T2*w) | Improved scan efficiency (multiple contrasts in one shot) and STN contrast with lower image distortion |
| Duchin et al. (2012) | 7T | 12 patients | TSE | STN | $0.39\times0.39\times2$ mm$^3$ | 5:30 min | No contrast assessments | 7 T MRI has comparable distortion in central region to 1.5T MRI, and can be used for DBS surgery |
| Deistung et al. (2013) | 7T | 9 healthy | Multi-orientation GRE (QSM) | STN, GPi & Vim | $0.4\times0.4\times0.4$ mm$^3$ | 16:57 min | Visual assessments for nuclei contrasts | High resolution QSM offers superb contrast for subcortical structures for the STN, GPi & Vim |
| Nowacki et al. (2015) | 3T | 13 patients | MDEFT | GPi | $1\times1\times1$ mm$^3$ | 12:00 min | MDEFT-based targeting offers 1.07mm error from the actual lead position | Direct targeting of GPi based on MDEFT is accurate and reliable. |

**Table S2**. Automatic segmentation algorithms for subthalamic nucleus (STN).

| **Authors** | **Field strength** | **No. of Subjects** | **Image contrasts** | **Resolution** | **Algorithm** | **Dice score** |
|---|---|---|---|---|---|---|
| Xiao et al. (2012) | 3T | 3 healthy & 3 patients | T2w & T2*w | 0.95×0.95×0.95 $mm^3$ | Multi-contrast label propagation | L: 0.63±0.06; R: 0.61±0.06 |
| Haegelen, et al. (2013) | 3T | 10 patients | T1w | 1×1×1 $mm^3$ | label propagation (ANIMAL registration) | L: 0.641±0.004; R: 0.640±0.006 |
| Haegelen, et al. (2013) | 3T | 10 patients | T1w | 1×1×1 $mm^3$ | label propagation (SyN registration) | L: 0.626±0.003; R: 0.640±0.006 |
| Haegelen, et al. (2013) | 3T | 10 patients (leave-one-out) | T2w | 1×1×1 $mm^3$ | Nonlocal means patched-based label fusion | L: 0.631±0.005; R: 0.575±0.022 |
| Xiao et al. (2014) | 3T | 10 patients (leave-one-out) | T2w | 1×1×1 $mm^3$ | Fuzzy label-fusion method | L: 0.74±0.08; R: 0.77±0.05 |
| Xiao et al. (2015) | 3T | 10 patients (leave-one-out) | T2*w & phase | 0.95×0.95×0.95 $mm^3$ | Multi-contrast nonlocal means label-fusion method | L: 0.877±0.026; R: 0.900±0.048 |
| Kim et al. (2015) | 1.5T & 7T | 26 patients (16 training & 10 testing) | T1w & T2w | Not provided | Feature-based shape prediction to transfer knowledge from 7T to 1.5T | L: 0.624±0.002; R: 0.659±0.011 |
| Visser et al. (2016) | 3T & 7T | 53 healthy | T1w, T2w, T2*w, QSM, FA | 0.5×0.5×0.5 $mm^3$ | Multi-contrast intensity-based generative model | ~0.8 |
| Park et al. (2019) | 3T | 102 patients (data augmentation added, 80 for training) | T2*w | 0.375×0.375×0.375 $mm^3$ | Deep convolutional neural network | 0.902 |
| Manjon et al. (2020) | 3T | 15 healthy (leave-two-out) | T2w | 0.5×0.5×0.5 $mm^3$ | Multi-atlas patch-based label fusion with multi-scales and multi-features | L:0.856; R: 0.855 |

**Table S3**. Survey of functional targeting methods for deep brain stimulation in treating Parkinson's disease (sMRI=structural MRI, DWI=diffusion-weighted imaging, fMRI= functional MRI).

| Author & Year | Target | Method | No. of subjects | Data modality | Reference space | Outcomes | Ground truth |
|---|---|---|---|---|---|---|---|
| Guo et al. (2005) | STN | Construct a functional atlas that contain electrophysiological data | 131 patients for atlas construction & 8 patients for testing | sMRI, physiological recording | CJH-27 atlas | 2.84 mm (Absolute differences between the centroids of the segmented STN and the real targets) | Real target location |
| D'Haese et al. (2005) | STN | Automatic selection of DBS target points using multiple electrophysiological atlases | 14 patients | sMRI and electrophysiologica l recording | Schaltenbrand-Wharen atlas | Mean deviation from final implant: 1.66 mm (Left) and 1.74 mm (right) | final implantation position in post-op CT |
| Nowinski et al. (2007) | STN | Construct a probabilistic functional atlas that represents spatial frequency of best contacts | 184 patients | sMRI, Physiological recordings | AC-PC based cartesian space | The "hot" functional STN resides inside anatomical STN | Coordinates in Schltenbrand-Wahren atlas |
| Sammartino et al. (2016) | Vim | Tractography-based methodology for the stereotactic targeting of the ventral intermediate nucleus | 14 patients and 15 healthy subjects | DWI | subject native space | Euclidean distance between tractography-based and surgical target was 1.6+/-1.1mm | Intra-operative stimulation & landmark-based targeting |
| Pallavaram et al. (2008) | STN | A probabilistic efficacy map was constructed with a spherical shell kernel | 19 patients (95 stimulation data) | T1w MRI, and clinical assessments | In-house atlas space | Final stimulation locations are within the "hot zones" of the efficacy map | final implantation position in post-op CT |
| Anderson et al. (2011) | Vim | A finger movement task is used to trigger activation in relevant regions | 58 healthy subjects | fMRI | MNI space | Within 3 mm of clinically optimized targets | Optimized targets based on anatomical landmarks |
| Horn et al. (2017) | STN | Linear regression based on functional and structural connectomic patterns is used to predict motor symptom improvements | 51 patients for training and 44 patients for testing | fMRI & DWI | MNI space | An average error of 15% UPDRS improvement | Post-surgical UPDRS motor scores |

Supplementary material

| | | | | | | | |
|---|---|---|---|---|---|---|---|
| Akam et al. (2017) | STN | Voxel-based statistical analysis was used to identify significant treatment clusters, which was mapped to relevant cortical regions | 20 patients | T1w MRI & DWI | MNI space | The maximum overall efficacy is X=-10(-9.5) mm, Y=-13(-1) mm and Z=-7(-3) mm in MNI(AC-PC) space. | functional outcomes before and 1-yr after surgery |
| Baumgarten et al. (2018) | STN | An artificial neural network is trained based on stimulation loci coordinates and electric current to predict therapeutic benefits and side effects | 30 patients (130 tested electrode positions) | T1w MRI and clinical assessments | ParkMedAtlas atlas | The mean sensitivity and specificity were 93.07% and 69.24% for the therapeutic effect, 73.47% and 91.82% for the side effect | Therapeutic and side effects (no effect, partial effect, or full effect) 3 months before and after surgery |
| Middlebrooks et al. (2018) | GPi | 10 segments of GPi is derived from structural connectivity to 10 cortical and subcortical structures; activation of each segment is correlated with motor symptom improvements | 11 patients | T1w MRI & DWI | Subject native space | GPi segments related to primary and supplementary motor areas are strongly corrected with UPDRSIII improvement | UPDRSIII scores from preoperative and 6-month follow-up visits |
| Coenen et al, (2020) | DRT | DRT was tracked and used as a direct thalamic or subthalamic target | 36 patients | T1w & T2w MRI, DWI | Subject native space | Close distance to the DRT is correlated with tremor reduction | Surgical outcomes |
| Lin et al. (2020) | STN | Random forest classifier based on mean FA and streamline fraction | 58 patients for training, 19 for testing | T1w & T2w MRI, DWI | Subject native space | 84.9% accuracy to discriminate ineffective targets | Surgical outcomes |

**Table S4**. Public surgical navigation software for deep brain stimulation (sMRI=structural MRI, DWI=diffusion-weighted imaging).

| **Name** | **Platform** | **Imaging modality** | **Trajectory planning** | **Electrode reconstruction** | **targeting technique** | **Preinstalled atlas** |
|---|---|---|---|---|---|---|
| CranialVault (D'Haese et al., 2012) | CRAVE | CT & sMRI | Yes | Yes | Based on probabilistic functional atlas | In-house electrophysiological and subcortical atlases |
| PyDBS (D'Albis et al., 2015) | 3D Slicer | sMRI | yes | No | Based on structural atlas | In-house population-averaged 3T MRI atlases (T1w & sectional T2w) of PD patients |
| DBSproc (Lauro et al., 2016) | AFNI | sMRI, DWI & CT | No | Yes | Patient-specific direct targeting | N/A |
| IBIS (Drouin et al., 2017) | IBIS neuronav | sMRI, angiography, CT | Yes | No | Based on structural atlas | N/A |
| Lead-DBS (Horn et al., 2018) | MATLAB | sMRI, tracrography | No | Yes | Based on structural and connectomic atlases | 17 cortical & 18 subcortical public atlases and averaged brain connectomic atlas |
| PaCER (Husch et al., 2018) | MATLAB | sMRI & CT | No | Yes | NA | N/A |